\documentstyle[aaspp4]{article}
\begin{document}
\title{Emission spectrum of Sagittarius A$^*$ and the  neutrino ball scenario} 
\author{Faustin Munyaneza and Raoul D. Viollier}
\affil{Institute of Theoretical Physics and Astrophysics,
 University of Cape Town, Rondebosch 7701,\\
Republic of South Africa}
\begin{abstract}
The  emission spectrum of the supermassive compact dark object at the Galactic center
is calculated in the framework of standard thin accretion disk theory assuming
that the compact object  is a neutrino ball 
of $2.6 \times 10^{6}M_{\odot}$
instead of  a  supermassive black hole. The  neutrino
ball scenario  could explain 
the observed radio and infrared emission spectrum of the Galactic center
for wavelengths between $\lambda=0.3 \ {\rm cm}$ and
$\lambda=10^{-3} \ {\rm cm}$, if the neutrino mass and the accretion rate
fulfil some constraints.
\end{abstract}
\section{Introduction}
It is generally accepted that
accretion onto
compact objects is the most efficient mechanism
of transforming gravitational potential energy
into radiation (see, e.g., Frank et al. 1992).
Sagittarius A$^*$ (Sgr A$^*$ ) at the Galactic center is  an unusual
source of radiation which has remained  a longstanding mystery.
The dynamics of stars  around the Galactic center 
(Eckart and Genzel 1996, 1997; Genzel et al. 1996,1997 and 
Ghez et al. 1998) is usually interpreted as evidence for
 a supermassive black hole
 of mass $\sim 2.6 \times 10^{6}M_{\odot}$ near Sgr A$^*$.
Observations of gas flows in the vicinity 
of Sgr A$^*$ reveal a mass
accretion rate onto the central object of $\sim 10^{-4}M_{\odot}{\rm yr}^{-1}$
(Melia 1992; Genzel et al. 1994).
In  standard thin accretion  disk theory, with a reasonable efficiency
of  $\sim 10 \% $, this accretion rate would correspond
to a luminosity of $ \sim 10^{42}~{\rm erg \ s^{-1}}$.
However, the actual luminosity observed
is $\sim 10^{37} {\rm erg \  s^{-1}}$.
 Moreover, the spectrum
is essentially flat in  $\nu L_{\nu}$ from radio waves to X-rays,
with  the exception of a few bumps (Rogers et al. 1994,
Menten et al. 1997, Predehl and Tr\"{u}mper 1994,
Merc et al. 1996). Thus both the  observed low luminosity
and the spectral energy distribution differ
 very  much from the spectrum that would be
expected from a standard thin disk around a supermassive black hole.
This discrepancy is known as the  ``blackness problem''
of the Galactic center.
Both the blackness of Sgr A$^*$ and
its peculiar spectrum  were the  source of exhaustive debate
in the recent past.
Several models for the accretion and the emission spectrum 
of Sgr A$^*$ have been proposed.
Melia (1994) modelled the spectrum of  Sgr A$^*$ as   synchrotron
 radiation emitted by thermal electrons,
heated through the dissipation of magnetic energy,
as a result of a Bondi-Hoyle accretion
 process 
fed by winds emanating from stars in the vicinity  of Sgr A$^*$.
Optically thick synchrotron radiation emitted by
 a  jet-disk system was also proposed 
as an explanation for the radiation of Sgr A$^*$
(Falcke, Mannheim and Biermann 1993;
 Falcke  and Biermann 1996). Moreover,
 synchrotron radiation emitted by  
 a quasi-monoenergetic  ensemble of relativistic electrons
(e.g.  Beckert and Duschl 1997) has been put forward
as a possible
emission mechanism.

Probably the most sophisticated model that is consistent
with the observed emission spectrum of Sgr A$^*$  from radio waves
to $\gamma$ rays is based on   Advection Dominated Accretion Flows (ADAF)
(Narayan et al. 1995, 1998; Mahadevan 1998; Manmotto  et al. 1997). 
This model relies on the idea
that  most of the
 energy released by viscous dissipation 
is stored as thermal energy in the gas which is then  advected
to the center, thereby radiating off  only a small fraction of the energy
 (Narayan and Yi 1995; Abramowicz et al. 1995).
 An essential ingredient of the ADAF models is  that the compact dark
object at the
 Galactic center is a {\it black hole}.
In fact, the existence of an event  horizon around the black hole
 is essential  in order to ensure
that  whatever energy falls  into the central object
 disappears  without being re-radiated.
 This model also requires the protons  to have a much
higher temperature than the electrons,
and the gas must therefore have a two-temperature structure.
However, it has also been recently pointed out that 
ADAF models,  as a solution 
of  astrophysical accretion problems,
should be treated with some caution,
as their physical basis is somewhat uncertain 
 (Bisnovatyi-Kogan  and Lovelace 1999).
Moreover, it is important to note that
none of the above models, including ADAF, 
can  predict the intrinsic shape and size of Sgr A$^*$
as observed at 7 mm ( Lo et al. 1998 ). It is also worthwhile to note
that the theoretical models for the emission of Sgr A$^*$
are unable to
explain the VLBI observations  of Sgr A$^{*}$, revealing that
 the observed size follows a $\lambda ^{2}$ dependence
and  the apparent source structure can be described
by an elliptical Gaussian brightness distribution
(Davies et al. 1976; Lo et al. 1985, 1993; Rogers et al. 1994;
Krichbaum et al. 1998; Bower  and Backer 1998).

A direct proof of the existence of a supermassive
 black hole would require
the observation of objects that are moving at relativistic velocities at
distances close to
the Schwarzschild radius.
However,  the best current observations
only probe the gravitational potential
at radii $4\times 10^{4}$ larger than the Schwarzschild radius
of a black hole of mass $2.6 \times 10^{6} M_{\odot}$ (Ghez al. 1998).
Thus there is no compelling direct evidence for 
the existence of  a supermassive
black hole   at the Galactic center. It is therefore
perhaps prudent not to focus too much
on the black hole scenario as the only  possible solution
 for  the  supermassive compact
dark object at the Galactic center,
without having explored alternative scenarios.

For instance, a compact dark stellar cluster could be an
alternative to the black hole scenario.
However, such clusters obey stringent stability criteria
(see, e.g., Maoz 1995, 1998).
A viable cluster must thus  have both evaporation and collision time scales
larger
than the lifetime of our Galaxy, i.e. $\sim 10~ {\rm Gyr}$,
and this is  more likely to be fulfilled with a cluster of substellar
objects.
But, apart from a compact cluster of  very low-mass black holes 
or brown dwarfs that is
free of stability problems, the most attractive
alternative to a dense  and dark stellar  cluster is
a cluster of elementary weakly interacting particles.

In fact,  
 such  an alternative model for the supermassive
compact dark object at the Galactic center
has been developed (Viollier et al. 1992, 1993; Viollier 1994;
Tsiklauri and Viollier 1996, 1998a,b, 1999; Bili\'c et al. 1998;
Bili\'c and Viollier 1999a,b).  Tsiklauri and Viollier (1998a) have  argued
that  the Galactic center is made of nonbaryonic
dark matter in the form of massive neutrinos 
condensed  in a  supermassive neutrino ball
of $2.5 \times 10^{6}M_{\odot}$ in which the degeneracy pressure
of the neutrinos
balances their self-gravity.
A supermassive neutrino ball differs from a black hole of the same
mass 
mainly by the shallow gravitational potential inside the neutrino ball.
Such neutrino balls could have been formed in the early Universe
	during a first-order gravitational
phase transition (Bili\'c and Viollier 1997, 1998, 1999a,b).
It has been shown that
the dark matter 
observed through stellar motion
at the Galactic center
(Ghez et al. 1998) is consistent with a supermassive
neutrino ball  
 of mass of $2.6 \times 10^{6}$ solar masses made of self-gravitating
heavy neutrino matter (Munyaneza, Tsiklauri and Viollier 1999).
Moreover , it has been pointed out that tracking
the orbit of the fast moving star S1 (Genzel et al. 1997) or S0-1
(Ghez et al. 1998), which is perhaps moving
 inside the neutrino ball,  offers the possibility to distinguish, 
within a few years time, the supermassive black hole scenario
from that of the neutrino
ball, for  the compact dark object at the Galactic center
(Munyaneza, Tsiklauri and Viollier 1998, 1999).

 The purpose of this paper is to calculate
the spectrum 
of the compact dark object at the Galactic center
based on standard thin accretion disk theory,  assuming that
this object  is a supermassive neutrino ball rather than a black hole.
 We perform the calculation  of
the spectrum based on the most recent Ghez et al. 1998 data,
including the error bars of the observations.
While the observed motion of stars near the
Galactic center yields a lower limit for the neutrino mass $m_{\nu}$,
the observed infrared drop of the emission spectrum of Sgr A$^*$ provides
us with an upper limit for $m_{\nu}$.
A distance to the Galactic center of 8 kpc has been assumed throughout
this paper. 
The outline of this paper is as as follows: In section~2
we present the formalism used to calculate the spectrum  in the
 neutrino ball   scenario, and in section~3 we summarize and discuss
our results.

\section{ Model and results}
The basic equations which govern the structure of
neutrino balls have been derived in
a series of papers (Viollier et al. 1992, Viollier et al. 1993,
Viollier 1994, Viollier and Tsiklauri 1996, Bili\'c  and Viollier 1999a,b);
we thus can be very brief here.
Let us denote  the dimensionless  neutrino Fermi momentum  by
 $X=p_{\nu}/(m_{\nu} c)$, where $p_{\nu}$
 stands for the 
local Fermi momentum of the neutrinos of mass $m_{\nu}$.
The structure of the neutrino ball is governed by a system 
of two coupled
 differential
equations (Bili\'c, Munyaneza and  Viollier 1999), i.e.
\begin{equation}
\frac{dX}{dx}=-\frac{\mu}{x^{2}X} \ , 
\label{eq:01}
\end{equation}
\begin{equation}
\frac{d\mu}{dx}=\frac{8}{3}x^{2}X^{3} \ ,
\label{eq:02}
\end{equation}
subject to the boundary condition
$X(0)=X_{0}$ and $\mu(0)=0$.
In Eqs.~(\ref{eq:01}) and (\ref{eq:02}), $x$ stands for the 
dimensionless radial coordinate $x=r/a_{\nu}$, $\mu$ 
denotes the dimensionless mass
enclosed within a radius $x$, i.e. $\mu=m(r)/b_{\nu}$, and
 $a_{\nu}$ and $b_{\nu}$ are the length and mass scales, respectively,
which can be expressed as 
\begin{equation}
a_{\nu}=
2 \sqrt{\frac{\pi}{g_{\nu}}}\left(
\frac{M_{\rm Pl}}{m_{\nu}}\right)^2 L_{\rm Pl}
=2.88233\times10^{10}g_{\nu}^{-1/2}
\left(\frac{17.2\ {\rm keV}}{m_{\nu}c^{2}}\right)^{2}{\rm km},
\label{eq:03}
\end{equation}
\begin{equation}
b_{\nu}=
2 \sqrt{\frac{\pi}{g_{\nu}}}\left(
\frac{M_{\rm Pl}}{m_{\nu}}\right)^2 M_{\rm Pl}
=1.95197\times10^{10}M_{\odot}g_{\nu}^{-1/2}
\left(\frac{17.2\ {\rm keV}}{m_{\nu}c^{2}}\right)^{2} ,
\label{eq:04}
\end{equation}
in terms of Planck's mass and length, $M_{\rm{Pl}}=(\hbar c/G)^{1/2}$
 and  $L_{\rm{Pl}}=(\hbar G/c^3)^{1/2}$, respectively.
Here, 
 $g_{\nu}$ is the spin degeneracy
factor of the neutrinos and antineutrinos, i.e. $g_{\nu}=2$
for Majorana and $g_{\nu}=4$ for Dirac neutrinos and antineutrinos.
By choosing the appropriate Fermi momentum  and thus the neutrino
density ( $\sim X^{3}$) at the center
of the neutrino ball, we can
construct a solution  corresponding to a neutrino ball of
 $2.6\times 10^{6}M_{\odot}$.
In order to describe the compact dark object at the Galactic
center as a neutrino ball, and constrain
its physical parameters appropriately, it is worthwhile
to use the most recent
 observational data by Ghez et al. 1998, who established that
the mass enclosed within 0.015 pc at the Galactic center is
$(2.6 \pm 0.2) \times 10^{6} M_{\odot}$ solar masses.
Following the analysis of Tsiklauri and Viollier 1998a, 
the constraints 
on the neutrino mass 
 $m_{\nu}$, in order  to reproduce the observed  matter
distribution (Munyaneza, Tsiklauri \&  Viollier 1999), are for
 a  $M=2.4\times 10^{6} M_{\odot}$
neutrino ball  $m_{\nu} \ge 20.81~{\rm keV}~g_{\nu}^{-1/4}$,
and the radius of the neutrino ball  therefore 
obeys $R \le 1.50\times 10^{-2}~{\rm pc} $.
Using the value  $M=2.6 \times 10^{6}M_{\odot}$, the bounds on the 
neutrino mass are $m_{\nu} \ge 18.93~{\rm keV}~g_{\nu}^{-1/4}$, 
and the radius  of the neutrino 
ball  turns out to be $R \le 1.88 \times 10^{-2}~{\rm pc}$.
Finally,  for a  $M=2.8 \times 10^{6}M_{\odot}$ neutrino ball, 
the range of the
neutrino mass
is $m_{\nu} \ge 18.21~{\rm keV}~g_{\nu}^{-1/4}$, and the corresponding
neutrino ball radius $R \le 2.04 \times 10^{-2} {\rm pc}$.
We can calculate the angular velocity $\Omega$ of the 
matter falling onto the neutrino ball as
\begin{equation}
\Omega=\sqrt{\frac{Gm(r)}{r^{3}}}=\frac{c}{a_{\nu}}\sqrt{\frac{\mu}{x^{3}}},
\label{eq:05}
\end{equation}
where $G$ is Newton's gravitational constant.
The total mass of the neutrino ball
is $M=m(R)$ .
In the case of a black hole, we have  $M=m(r)$ 
already for radii much larger than the Schwarzschild radius.
In Fig.~1, we plot
the angular velocity as 
a function of the distance from the center
for a neutrino ball of mass 
$M=2.6 \times 10^{6} M_{\odot}$
and a neutrino mass  $m_{\nu}g_{\nu}^{1/4}c^{2}=18.93~{\rm keV}$.
The angular velocity corresponding to a black hole
of the same mass is also shown for comparison.
Close to the  center of the neutrino
ball, $\Omega(r)$ is nearly constant,
and the mass enclosed within a radius $r$
therefore  scales as $r^{3}$.

In the standard theory of steady and
geometrically thin accretion disks,
the power liberated in the disk per unit area is given
by (Perry \& Williams 1993; Frank et al. 1992)
\begin{equation}
D(r)=-\frac{\dot{M} \Omega(r) \Omega '(r) r}{4 \pi}
\left[1-\left(\frac{R_{i}}{r}\right)^{2}
\left(\frac{\Omega_{i}}{\Omega}\right)\right] \ .
\label{eq:06}
\end{equation}
Here
$R_{i}$ is the inner radius of the disk and $\Omega_{i}$
defines the angular velocity
 at the radius where  $\Omega(r)$ has a maximum, i.e.
$\Omega_{i}=\Omega(R_{i})$.
The prime on the function $\Omega(r)$ denotes the
derivative with respect to $r$.
The accretion rate $\dot{M}$
is  parametrized  as
\begin{equation}
\dot{M}= \dot{m}  \dot{M}_{{\rm Edd}} \ ,
\end{equation}
where $\dot{M}_{{\rm Edd}}=2.21\times 10^{-8}M \ {\rm yr}^{-1}$
denotes
 the Eddington limit accretion rate.
The maximal and minimal accretion rate allowed by the observations
are
  $\dot{m}=4\times 10^{-3}$ and $10^{-4}$
(Narayan et al. 1998), respectively.
The outer radius of the disk  has been  taken as 
$10^{5}$ Schwarzschild radii, since for larger radii,
the disk is unstable against self-gravity (Narayan et al. 1998).
We now use Stefan-Boltzmann's law, assuming that
 the gravitational binding energy
is immediately radiated away
\begin{equation}
D(r)=\sigma T^{4}_{{\rm eff}}(r) \  ,
\label{eq:07}
\end{equation}
where $\sigma$ is the Stefan-Boltzmann constant. 
The effective temperature $T_{{\rm eff}}$ can be easily derived using 
 Eqs. (\ref{eq:05}), (\ref{eq:06}) and (\ref{eq:07}) yielding
\begin{eqnarray}
T_{{\rm eff}}(r)&=&\left(\frac{\dot{M}_{{\rm Edd}}G}{8\pi\sigma}\right)^{1/4} 
\frac{b_{\nu}^{1/4}}{a_{\nu}^{3/4}}
 \dot{m}^{1/4} 
\left(\frac{3\mu-\mu 'x}{x^{3}}\right)^{1/4}
 \left[1-\left(\frac{x_{i}}{x}\right)^{2}
\frac{\Omega_{i}}{\Omega}\right]^{1/4} \nonumber \\
& = & T_{0}  \dot{m}^{1/4} 
\left(\frac{3\mu-\mu 'x}{x^{3}}\right)^{1/4}
 \left[1-\left(\frac{x_{i}}{x}\right)^{2}
\frac{\Omega_{i}}{\Omega}\right]^{1/4} \ ,
\label{eq:08}
\end{eqnarray}
where the constant $T_{0}$ is given
by 
\begin{equation}
T_{0}=\left(\frac{\dot{M}_{{\rm Edd}}G}{8\pi\sigma}\right)^{1/4} 
\frac{b_{\nu}^{1/4}}{a_{\nu}^{3/4}} \ .
\label{eq:09}
\end{equation}
Once the temperature  distribution in the disk is specified,
one can find its luminosity at a frequency $\nu$ using
\begin{equation}
\frac{dL_{\nu}}{dr}=\frac{16\pi^{2}h \nu^{3}\cos(i)}{c^{2}} \frac{r}
{\exp\left(\frac{h \nu}{k_{B}T_{{\rm eff}}}\right) -1} \ ,
\label{eq:10}
\end{equation}
with $L_{\nu}(x_{i})=0$.
In   Eq.~ (\ref{eq:10}), $h$  and $k_{B}$ are   Planck's and
Boltzmann's constants, respectively,
and  the disk inclination angle
 $i$
  is assumed to be $60^{0}$ as in Narayan et al. (1998). 
Picking up a particular value for $\nu$, we may integrate Eq. (\ref{eq:10})
numerically, taking the inner radius of the disk  to be determined by
$\Omega'(r)=0$.
 However , the inner radius of the accreting 
disk can be chosen to be zero, as the inner region,
where  $\Omega(r)$ is nearly constant,
 does not contribute
to the emission spectrum anyway .
It is worthwhile to note, that in the case
of a neutrino ball, there is no last
stable orbit, in contrast to the black hole case,
where the inner radius of the disk 
is taken to be three Schwarzschild radii.
The results of this integration are shown in
Figure~2,
where  the spectrum emitted  in the case of accretion
onto a black hole (dotted lines) of
 $M=2.6\times 10^{6}M_{\odot}$ is shown as well.
Here,  accretion rates of 
 $\dot{m}=10^{-3}$, \  $10^{-4}$ and $10^{-9}$ have been assumed
 for both scenarios.
Also shown in  
this plot are  the most up-to-date observations
of the emission spectrum of the  Galactic center (Narayan et al. 1998).
 The arrows represent  upper limits,
 and the box at a frequency  $\sim 10^{17}~{\rm Hz}$ stands for
 the uncertainty
in  the observed X-ray flux.
The open and filled squares represent  various
flux measurements and upper limits for Sgr A$^*$.
The open squares stand for the low angular resolution points
while the filled squares represent  the data points
with best resolution.
The observed spectrum rises at radio
 and submillimeter frequencies
of $\nu \simeq 10^{9}$ to $10^{12} {\rm Hz}$, where most of the 
emission occurs, and it  has a
sharp drop in the infrared. The X-ray 
observations consist of a possible
detection at soft  X-ray energies, and 
firm upper limits in the hard X-rays.
As seen in Fig.~2, the neutrino ball model reproduces the observed
spectrum from the  radio ($\lambda=0.3 \ {\rm cm}$)
 to the near infrared band ($\lambda=10^{-3} \ {\rm cm}$) very well.
Thus, as our model fulfils two of the most stringent conditions, i.e.
it is consistent  with  the
mass
distribution (Genzel et al. 1997, Ghez et al. 1998) and the bulk  part of  the emitted
spectrum,
we  conclude
that the neutrino ball
scenario is not in contradiction with most of  the 
currently available observational 
data.
As we see from Fig.~2  and also
as pointed out by Narayan et al. 1998,
the curves corresponding to the black hole (lines~ 4, 5 and 6) provide a
poor fit to the observational data.
A starving black hole , with  an accretion rate of $\dot{m}=10^{-9}$
(line~6 in Fig.~2)
would not fit the observed spectrum either.
This is in fact the main  reason why 
  standard accretion disk  theory was 
abandoned   as  a possible candidate
 for the description of the  Sgr A$^{*}$ spectrum (Narayan et al.  1995).
 Figure~3 shows the temperature
of the  disk as a function of the radius,
for an accretion rate of $\dot{m}=10^{-3}$ in  both scenarios.

The spectrum presented in Fig.~2
 corresponds to a neutrino ball or black hole
of $M=2.6 \times 10^{6}M_{\odot}$. 
To draw definite conclusions about the emission  spectrum of a neutrino
ball, 
it is necessary 
to investigate the dependence
of the spectrum
on i) the  mass of the neutrino ball ;
ii) the neutrino mass $m_{\nu}$, both with the ranges
allowed
by the  Ghez et al. 1998 data.
In Fig.~4, we present the  emission spectrum 
for a variety of neutrino ball masses, i.e.
 $M=2.4, 2.6, 2.8 \times 10^{6}M_{\odot}$.
From this plot, we conclude that, within the uncertainties,
 the  mass of the neutrino ball
has no significant effect on the spectrum of the neutrino ball.
In Fig.~5, we plot the spectrum
as a function of the neutrino mass for different accretion rates.
The top panel represents the spectrum for
an accretion rate of  $\dot{m}=10^{-4}$  while
the lower describes  an accretion rate of  $\dot{m}=10^{-3}$. 
The neutrino mass $m_{\nu}$  has been varied
as shown on the plot.
As the observed emission spectrum
has a sharp drop 
in the infrared region, 
we  require 
the theoretical spectrum 
not to extend to frequencies beyond  the innermost data points
 of the infrared drop  of the observed  
spectrum, yielding an upper bound for the neutrino mass. 
For each value of the accretion rate,
an upper bound for the neutrino mass is  established using
this condition.
This is reflected in Fig.~6, where we plot
the neutrino mass $m_{\nu}c^{2}$
as a function of the accretion rate  $\dot{m}$.
The vertical arrows pointing down show the inferred upper limits
of the neutrino mass for each accretion
rate. Thus
for $\dot{m} =10^{-4}$, the upper limit
 is $m_{\nu}c^{2}g_{\nu}^{1/4} \leq 29.73~{\rm keV}$;
for $\dot{m}=8\times10^{-4}$, the range of the neutrino mass
 is $m_{\nu}c^{2}g_{\nu}^{1/4} \leq 19.74~ {\rm keV}$;
for $\dot{m} =10^{-3}$, the neutrino mass is constrained by
 $m_{\nu}c^{2}g_{\nu}^{1/4} \leq 18.93~ {\rm keV}$;
and finally for
$\dot{m} =4\times10^{-3}$, the upper limit 
is found to be $m_{\nu}c^{2}g_{\nu}^{1/4} \leq 17.24 ~ {\rm keV}$.
The horizontal line shows the lower limit
on the neutrino mass obtained
by fitting the mass distribution of the neutrino ball
with the currently best observational data (Ghez et al. 1998).
Combining both  upper  and  lower limits for the neutrino
mass, we  arrive at  the following constraints for the neutrino mass
\begin{equation}
18.93~{\rm keV} \leq m_{\nu}c^{2}g_{\nu}^{1/4} \leq 29.73~{\rm keV}
 \ {\rm for} \  \dot{m}=10^{-4} ,
\end{equation}
\begin{equation}
18.93~{\rm keV} \leq m_{\nu}c^{2}g_{\nu}^{1/4} \leq 19.74~{\rm keV}
\ {\rm for}  \ \dot{m}=8\times10^{-4} \ .
\end{equation}
From Fig.~6, we  may conclude:
 i)In order to be consistent with the observational
Ghez et al. 1998 data, the accretion rate $\dot{m}$ onto
the neutrino ball should be less than $\sim 10^{-3}$,
implying an accretion rate $\dot{M}$ onto the neutrino 
ball that is less than  $\sim 5.7 \times 10^{-5}M_{\odot}{\rm  yr}^{-1}$;
ii) The neutrino mass range  is bounded from below by
the Galactic kinematics and also
bounded from above  by the spectrum.
 The range of allowed  values of the neutrino mass
 narrows as the accretion rate 
increases, vanishing at $\dot{m} \ge 10^{-3}$.

\section{Summary and discussion}
We have studied the emission spectrum of Sgr A$^{*}$  assuming
that it is a  neutrino
 ball of mass $M=(2.6 \pm 0.2)\times 10^{6}M_{\odot}$
with a size of a few tens of light days.
We have shown that, in this case,
the theoretical
spectrum, calculated in standard thin accretion disk theory, fits
 the observations
in the radio and infrared region of the spectrum
much better in the neutrino ball than in
 the black hole scenario, as seen from Fig.~2.
This is because, in  the neutrino ball scenario,
 the accreting matter experiences a much
shallower gravitational potential than in the case of
a black hole  with the same mass, and therefore less
viscous torque will be exerted.
Here , we note that the emitting region
for this part  of the spectrum
is of the order of the size
of the neutrino ball, i.e. a few tens of light days.
We have shown that the error bars
in the  mass of the neutrino
ball have practically no significant
effect
on the spectrum of Sgr A$^*$.
By assuming that the emission  spectrum cannot
extend  beyond the observed innermost data  points of
 the infrared drop 
of the Sgr A$^*$ spectrum,
we have established
that the range of possible values of the neutrino mass
 narrows as the accretion
rate $\dot{m}$ increases.
We have also
shown that an accretion rate
of more than $\dot{M} > 5.7 \times 10^{-5}M_{\odot}{\rm yr}^{-1}$
would render the allowed  range of  neutrino masses
 inconsistent with the lower limit
obtained 
from the observational data based on the kinematics of stars.

The thin accretion disk  neutrino ball scenario alone  can,
 of course, neither
explain
the lower part of the radio spectrum, i.e.
 $\nu \stackrel{\textstyle <}{\sim} 2 \times 10^{11}~{\rm Hz}$,
nor  can it explain  a possible spectrum 
 for  $\nu  \stackrel{\textstyle >}{\sim} 10^{14}~{\rm Hz}$.
The latter is  a consequence of the fact that the escape velocity
from the center of the neutrino ball of
$2.6 \times 10^{6} M_{\odot}$ is only  about 1700 km/s.
In order to get X-rays , the particles
need to reach a sizable fraction
of the velocity of light, which is impossible in
the pure  neutrino ball scenario.
However, as the heavy neutrinos presumably  decay 
radiatively ($\nu_{\tau} \rightarrow \nu_{\mu}+\gamma$ or
$\nu_{\tau} \rightarrow \nu_{e}+\gamma$)
with  a lifetime of $\stackrel{\textstyle >} {\sim} 10^{18}  {\rm yr}$
(assuming Dirac neutrinos and the current limits for the mixing
 angles),
 there  will
be some X-ray emission of the
 order of $\stackrel {\textstyle <}{\sim} 10^{34} {\rm erg \  s^{-1}}$
at an energy of $m_{\nu}c^{2}/2$, which could
be presumably detected by the CHANDRA X-ray satellite. Moreover, if  both 
 neutrinos and antineutrinos
are present in the neutrino ball,
 annihilation ($\nu_{\tau}+\bar{\nu}_{\tau} \rightarrow \gamma + \gamma$) will
also contribute to the X-ray spectrum at an energy $m_{\nu}c^{2}$, 
 concentrated at the center of the neutrino ball,
albeit with a much smaller luminosity (Viollier 1994).
Furthermore,
it is worthwhile to speculate that  a neutron star
at the Galactic
center, surrounded by  a neutrino  halo of 
 $M=2.6 \times 10^{6}M_{\odot}$,  might   
 explain the  observed spectrum of Sgr A$^{*}$.
A similar idea was proposed long ago by Reynolds  and McKnee 1980, who
suggested that
the radio emission of Sgr A$^{*}$ could be due
 to  an otherwise unobservable   radio pulsar.
However, as
 the accretion rate  onto
the neutrino ball 
is of the order of $\dot{M}=10^{-5}M_{\odot} \ {\rm yr}^{-1}$,
i.e. three orders of magnitude
larger than the Eddington accretion
rate  of $\sim 10^{-8}M_{\odot} \ {\rm yr}^{-1}$  onto a neutron star,
much of the baryonic matter falling towards this neutron
star will have to   
 be expelled before reaching the neutron star surface.

\section{Acknowledgements}
One of us (F. Munyaneza) gratefully acknowledges  funding
from the Deutscher Akademischer Austauschdienst (DAAD) and the University
of Cape Town. This work is supported by the Foundation
for Fundamental Research (FFR). We also thank D. Tsiklauri for
useful comments.

\newpage

\centerline{Figure captions:}
Fig~1: The angular velocity as
a function of the distance from the center
for the neutrino ball and the black hole scenarios. The  neutrino ball
and the black hole have the same mass  $M=2.6\times 10^{6}M_{\odot}$.

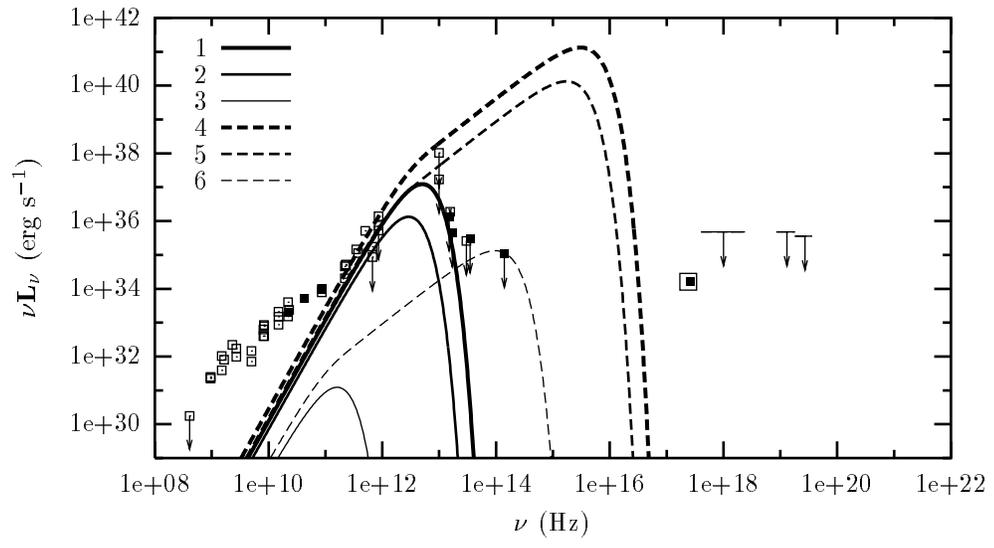
Fig~2: The spectrum of Sgr A$^*$ in both scenarios for various accretion
rates.
The continuous curves (lines~1,2,3) correspond to a disk
immersed in the potential of a neutrino ball
while the dashed lines (lines~4,5,6) correspond
to a disk around a black hole.
Lines 1 and 4 stand for  an accretion rate of  $\dot{m}=10^{-3}$,
while lines 2 and 5 correspond to an accretion rate of $\dot{m}=10^{-4}$.
 Finally,
 an accretion rate of $\dot{m}=10^{-9}$ for a starving disk
is represented by the  lines~3 and 6.
The observed data points, taken from Narayan et al. 1998, 
have been included in this plot.
The arrows denote   upper bounds.
The filled squares show the data with high resolution while 
the open circles represent the data with less resolution.

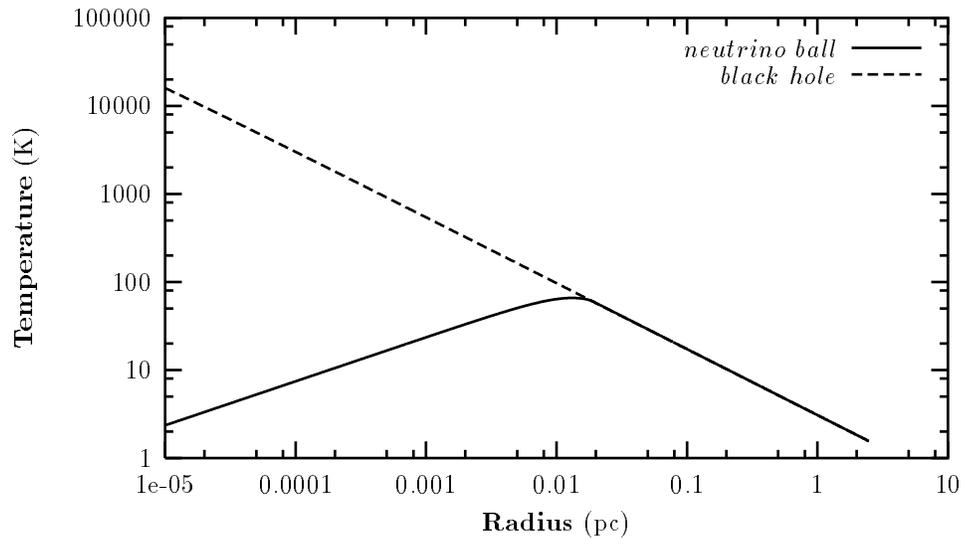
Fig.~3: The temperature of the disk as
a function of the distance from the center
for both scenarios. The accretion rate is $\dot{m}=10^{-3}$.

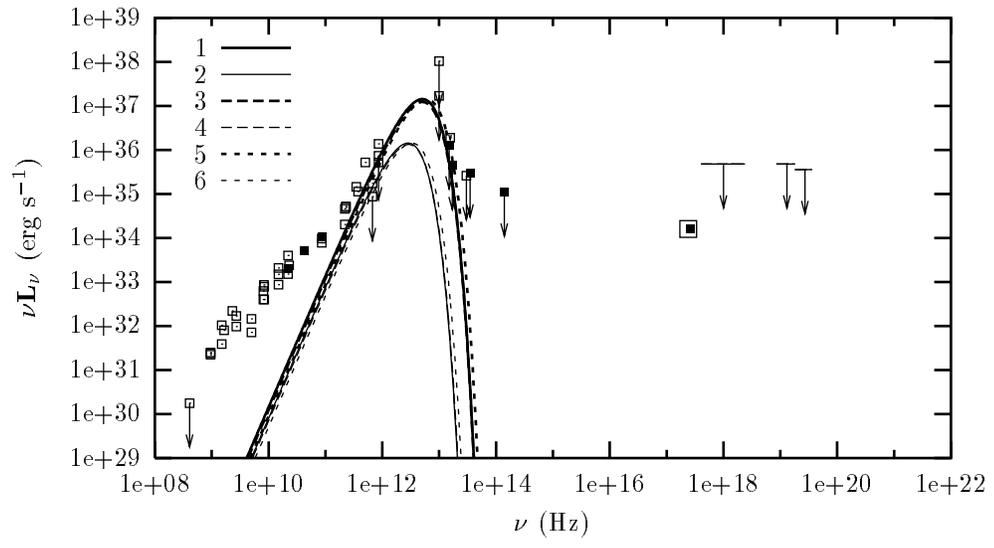
Fig.~4: The Sgr A$^*$ emission  spectrum for 
 neutrino ball masses  $M=2.4,2.6$ and $2.8 \times 10^{6}$ solar masses.
The thick lines (1,3,5) correspond to
an accretion rate of  $\dot{m}=10^{-3}$
while the thin lines (2,4,6) are drawn for $\dot{m}=10^{-4}$.
The  mass of the neutrino ball does not have
a significant effect on the spectrum of Sgr A$^*$.

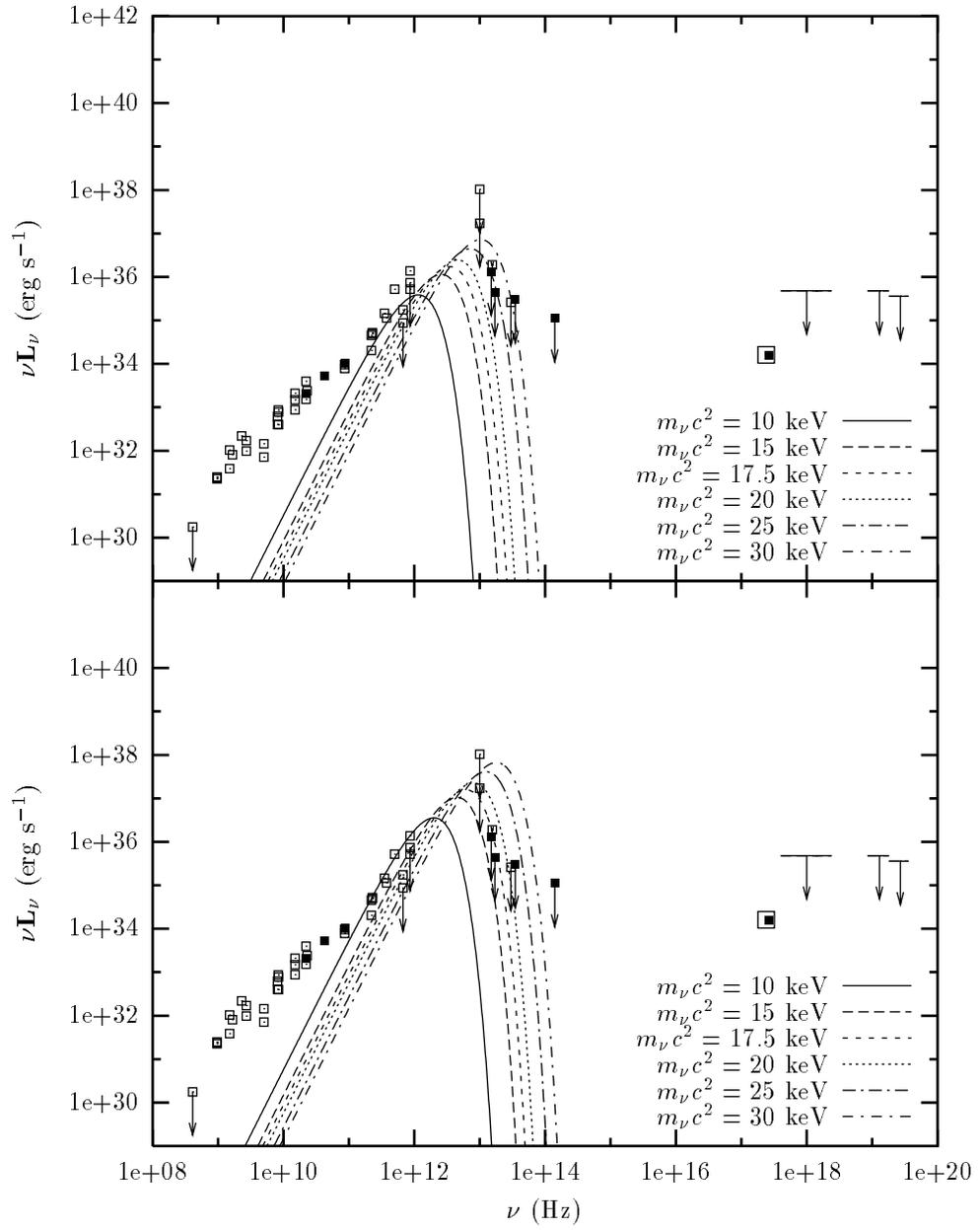
Fig.~5: The  Sgr A$^{*}$ emission spectrum  for various 
  neutrino masses $m_{\nu}$.
An upper limit for the neutrino mass
is inferred by requiring that the theoretical spectrum cannot
go beyond the innermost  points of the infrared drop of the 
observed spectrum. The top panel  represents the spectrum
for $\dot{m}=10^{-4}$ while the lower describes an accretion rate of
$\dot{m}=10^{-3}$.

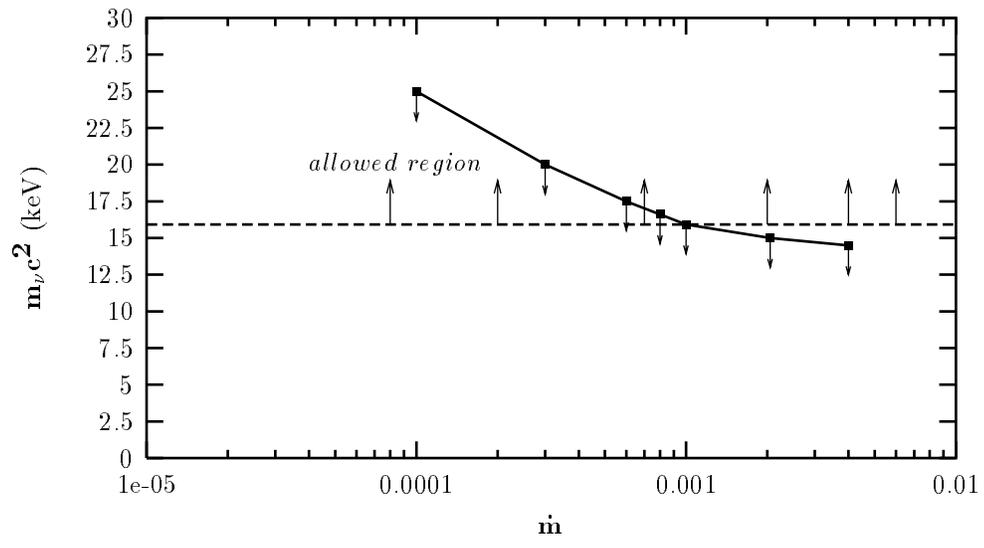
Fig.~6: The  neutrino mass $m_{\nu}$ as a function
of the accretion
rate $\dot{m}$ for $g_{\nu}=2$.
The horizontal line, with arrows pointing up, shows the lower limit 
of the neutrino mass, as obtained from the dynamics of stars.
The arrows pointing down denote the upper limit, 
determined from the drop  of the
spectrum in the
infrared region.
The range of the neutrino mass narrows as the accretion
rate $\dot{m}$ increases. For $\dot{m} > 10^{-3}$ ,
the upper limit on the neutrino mass becomes inconsistent
with the lower limit from the dynamics of stars.
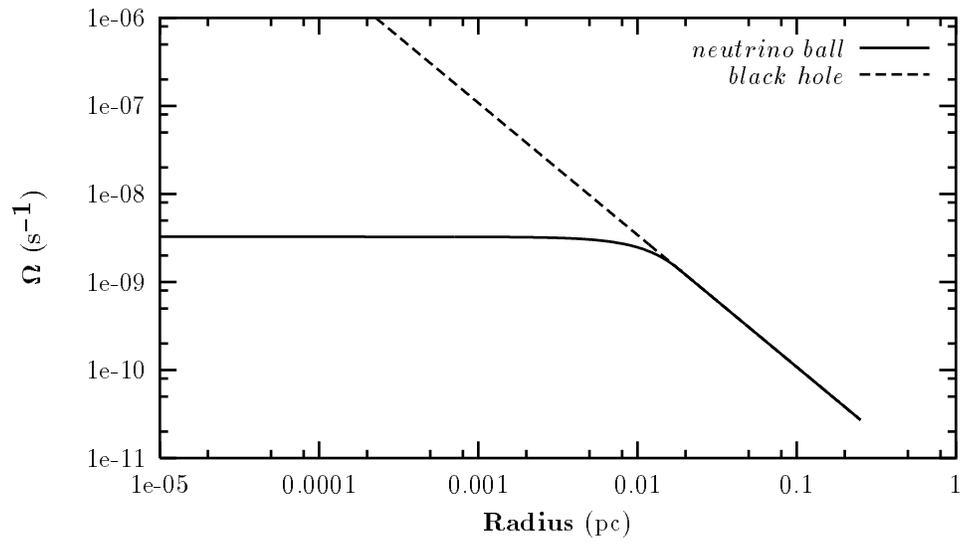
\end{document}